# Cooperative Highway Work Zone Merge Control based on Reinforcement Learning in A Connected and Automated Environment

Tianzhu Ren, Ph.D.
Amazon

Yuanchang Xie*, Ph.D., P.E.
Associate Professor
Department of Civil and Environmental Engineering
University of Massachusetts Lowell
1 University Ave, Lowell, MA 01854
Phone: (978) 934-3681
Email: Yuanchang_Xie@uml.edu

Liming Jiang, Ph.D. Student
Department of Civil and Environmental Engineering
University of Massachusetts Lowell
1 University Ave, Lowell, MA 01854
Email: Liming_Jiang@student.uml.edu

* Corresponding Author

Word Count: 4,953 + 250*1 (6 Figures and 1 Table) = 5,203 Words

## ABSTRACT

Given the aging infrastructure and the anticipated growing number of highway work zones in the United States, it is important to investigate work zone merge control, which is critical for improving work zone safety and capacity. This paper proposes and evaluates a novel highway work zone merge control strategy based on cooperative driving behavior enabled by artificial intelligence. The proposed method assumes that all vehicles are fully automated, connected and cooperative. It inserts two metering zones in the open lane to make space for merging vehicles in the closed lane. In addition, each vehicle in the closed lane learns how to optimally adjust its longitudinal position to find a safe gap in the open lane using an off-policy soft actor critic (SAC) reinforcement learning (RL) algorithm, considering the traffic conditions in its surrounding. The learning results are captured in convolutional neural networks and used to control individual vehicles in the testing phase. By adding the metering zones and taking the locations, speeds, and accelerations of surrounding vehicles into account, cooperation among vehicles is implicitly considered. This RL-based model is trained and evaluated using a microscopic traffic simulator. The results show that this cooperative RL-based merge control significantly outperforms popular strategies such as late merge and early merge in terms of both mobility and safety measures.

## INTRODUCTION

Bottlenecks generated by work zones as well as traffic incidents are one of the most important contributors to non-recurrent congestion and secondary accidents. Many previous work zone studies focused on merge control and proposed a variety of strategies such as early merge (EM)(*2*) and late merge (LM)(*3*) to improve work zone throughput. EM typically uses a sequence of "*DO NOT PASS*" signs that can be activated/deactivated depending on traffic to create a no passing zone of varying length. A traffic sensor is mounted on each sign to monitor traffic in the open lane. The purpose of the no passing zone is to encourage drivers in the closed lane to switch to the open lane before reaching the end of the dynamically changing queue (or slow-moving traffic) to improve safety and efficiency. EM often creates high-speed but low-density flow at the merging point. While for LM, drivers in both open and closed lanes are urged to stay in their respective lanes until the merging point, where they take turns to merge. Compared to EM, LM can effectively reduce the overall queue length, since both lanes are used for queue storage. However, LM often generates low-speed but high-density flow at the merging point. Ideally, the best merge control should result in high-speed and high-density flow.

Some advanced driving assistant systems such as Adaptive Cruise Control (ACC)(*4*) enable vehicles to drive at a high speed while maintaining a small gap (i.e., high density). Such a feature is only for improving vehicle longitudinal control and cannot address the challenging work zone merge problem. Also, an ACC-equipped vehicle only considers its interactions with the vehicle immediately in front of it and in the same lane (including vehicles attempting to merge into its lane), trying to make optimal decisions locally. To improve work zone traffic operations, it is important for individual vehicles to take global traffic conditions into consideration and cooperate with other vehicles in both the open and closed lanes.

To enable collaborative driving behavior among CAV, there are three main challenges: (a) how to effectively take the vast amount of unstructured traffic information into consideration; (b) how to choose an optimal control policy based on the dynamically changing surrounding traffic that maximizes the benefits of the subject vehicle in the long run instead of just the next few time steps; and (c) how to make the best decisions based on not only the subject merging vehicle's state, but also its surrounding vehicles' current states and possible moves in the future. Some previous studies have attempted to address the collaborative merge problem. Chen et al.(*12*) applied a gap acceptance algorithm and proposed several rules to decide a vehicle's actions before merging into the target lane. Urmson at al. (*11*) used a slot-based approach for cooperative merging control. These rule-based methods depend heavily on specific situations which are pragmatically vulnerable due to their inability to adapt to unforeseen environment.

Reinforcement learning (RL) has been successfully applied to a variety of fields with the growing availability of cost-effective high-performance computing hardware. RL together with deep neural networks can take large dimensions of state space into consideration, making it very appealing for work zone control. Using RL, analysts do not need to explicitly specify how a work zone changes from one state into another (i.e., state transition probability matrix), which dramatically reduces the modeling effort needed, particularly the trouble associated with specifying the uncertain state transition probability matrix. Vehicle agents can learn from a huge number of simulated scenarios about the complex nonlinear relationship between their next moves and work zone traffic operations, and find actions with the maximum long-term reward.

Due to these desirable features, RL has been applied in self-driving vehicles such as NVIDIA (*5*), Tesla Autopilot (*6*) and Google Waymo. Some researchers also applied RL in ramp metering (*13,14*). Specifically, Fares et al. (*13*) developed a RL model to optimally control the density of freeway mainstream for maximizing traffic throughput and minimizing travel time. Their model was formulated as a Markov Decision Process (*7*) and solved by Q-learning (*8*). Yang et al. (*14*) proposed a Deep Q-Network (DQN)(*9*) control strategy to identify the optimal ramp metering rate. The DQN considered upstream and downstream traffic volumes as the input state and chose either green or red for the ramp meter traffic light as the action at each decision interval. Yu et al. (*15*) applied deep Q-Learning to control a simulated car for turning and obstacle avoidance maneuvers. These studies all considered a discrete action space due to its simplicity and fast convergence, although many vehicle control problems (e.g., (*15*)) very likely may benefit more from using a continuous action space. Sallab et al. (*16*) compared a discrete action-space Deep Q-Network with a continuous action-space Deep Deterministic Actor Critic (DDAC)(*10*) for lane-keeping assistance based on an open source car simulator, and the results showed that the discrete DQN method led to abrupt steering maneuvers while the continuous DDAC method generated better performance and smoother control.

This research proposes a deep neural network based RL control approach that guides AVs through work zones. Specifically, a work zone is divided into two metering zones and a merging zone (see Figure 1). In the metering zones, AVs are not allowed to change lanes and they focus on adjusting longitudinal positions using the proposed RL method. By the time AVs reach the merging/lane reduction point, they will be able to maintain a sufficient front gap if all vehicles were projected onto a single lane. In this way, they can merge safely and form a high-speed and high-density vehicle platoon. The key to this proposed approach is how to adjust AVs' longitudinal positions properly in the metering zones. In this research, each AV in the closed lane is considered as a RL agent. It learns the best control strategy through its interactions with the simulated traffic environment using VISSIM. At each time step, this agent takes an action (i.e., acceleration, deceleration). At the next time step, the value for its previous action is updated based on a set of reward functions and the interactions between the agent and the environment. To improve the control model's generalization ability, a deep neural network is used to store the learning results. The proposed RL approach is detailed in the next section.

## METHODOLOGY

### Overview

As shown in Figure 1, a work zone is divided into two metering zones followed by a merging zone. All vehicles approaching the work zone are instructed to increase their distance headways upon entering Metering Zone I. Specifically, each vehicle needs to increase its front distance headway to twice the safe distance needed for the corresponding speed (assuming 70km/h). Metering Zone I is to provide sufficient distance (i.e., reaction time) for vehicles to double their front gaps and lane changing is prohibited in this zone. In Metering Zone II, vehicles in the open lane (left lane in Figure 1) will adopt the same car-following behavior as in Metering Zone I, while vehicles in the closed lane (right lane in Figure 1) are required to adjust their longitudinal positions. By the time vehicles reach the merging/lane reduction point, they will be able to maintain a sufficient front gap if all vehicles were projected onto a single lane. Following this longitudinal control strategy, towards the end of Metering Zone II, if vehicles in both lanes are projected onto a single virtual lane, all the distance headways are expected to be close to but greater than the minimum safe distance gap. In the Merging Zone, lane changes are allowed and vehicles in the two lanes take turns to merge. In summary, the core of the RL-based method is the longitudinal control in the two metering zones, where lane changes are prohibited. Before Metering Zone I, vehicles follow normal driving behavior. After Metering Zone II, vehicles also follow normal driving behavior other than being instructed to merge in the merging zone.

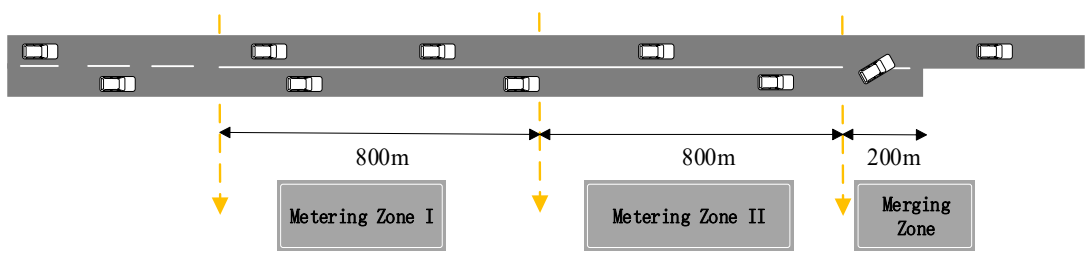

Figure 1 Overview of RL-based Control.

In this study, the deep neural network based RL strategy and other benchmark strategies are all evaluated using VISSIM microscopic traffic simulation. In Metering Zone II, vehicles in the right lane are controlled by a convolutional neural network trained by RL, and left-lane vehicles are controlled by a modified VISSIM default driving behavior model. The modification simply doubles the default time headway to create sufficient gaps for right-lane vehicles to merge in the Merging Zone.

**Deep Reinforcement Learning**

In this section, a detailed description of the RL approach is provided, including reinforcement learning basics, state representation, neural network architecture and soft actor critic (SAC) (*1*) RL and reward shaping.

In this research, the control of right-lane vehicles (see Figure 1) is formulated as a Markov Decision Process (MDP) consisting of numerous state $s$ primarily defined by the surrounding traffic. Based on the learned policy $\pi$, an action $a$ is selected at each state and executed. After the execution, the system (i.e., work zone traffic operations) will react to the action, from which a reward $r$ can be observed, and transit to a new state $s'$. The reward and the current and new states are then used to update the policy. To take each action's long-term reward into consideration, the expected discounted cumulative reward $\sum R$ is calculated along with the policy from the initial state (a vehicle enters the work zone) to the terminal state (a vehicle merges into the open lane in the merging zone).

There are two types of widely used reinforcement learning methods. The first type is model-free RL such as Q-learning and Sarsa. They try to find the optimal policy by updating a table that saves the long-term rewards for each state and action pair. They then use this table directly for identifying the best actions. The other type is model-based RL like DQN, Deep Deterministic Policy Gradient (DDPG) (*10*) and the proposed SAC, which learn a model to approximate the MDP and use this learned model to find the optimal control actions.

*State Representation*

In this study, the system state is defined by three components: network speed grid map, network acceleration grid map and an 8-element vector representing the traffic surrounding the subject vehicle being controlled by RL (e.g., the red vehicle in Figure 2).

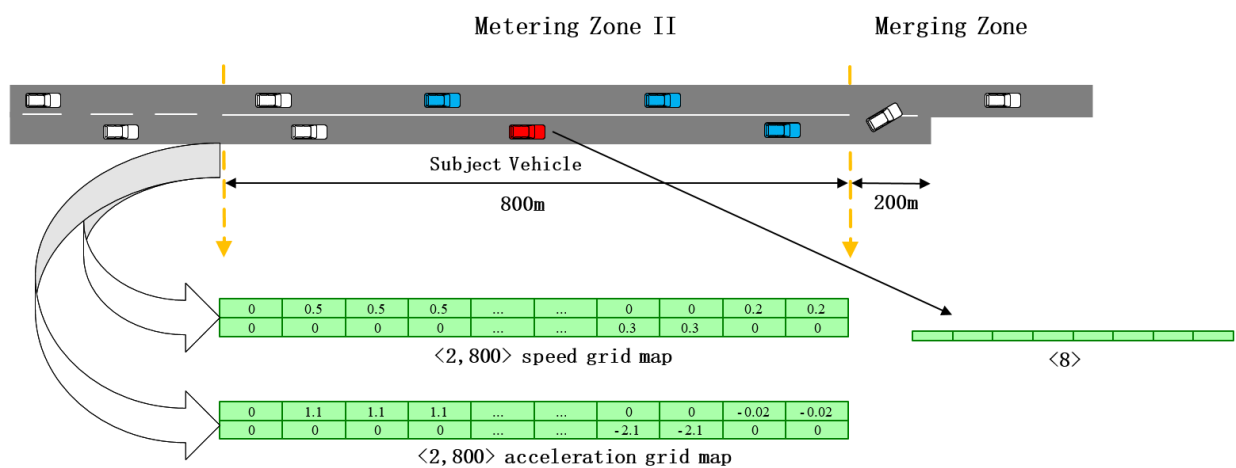

Figure 2 State representation

As in Figure 2, the 800-meter Metering Zone II is divided into 2 x 800 cells for the network speed grid map and network acceleration grid map. Each row is for a lane and each cell is for a 1-meter segment. The numbers in each cell represent either the speed or the acceleration of the vehicle occupying that cell. If a vehicle occupies multiple cells, the speed/acceleration values in the corresponding cells will be equal.

The speed values illustrated in Figure 2 are normalized based on the actual vehicle speeds and are bounded by 0 and 1. The normalization is done via dividing the original speed values by the maximum speed in the training and testing processes. Similarly, the acceleration values in Figure 2 are normalized using the maximum absolute value and are bounded by -1 and 1.

In addition to the speed and acceleration grid maps, an 8-element vector is included, which consists of: (a) the relative positions and relative speeds between the subject vehicle (e.g., the red vehicle in Figure 2) and some of its neighboring vehicles (blue vehicles in Figure 2), and (b) speed and position of the subject vehicle. As can be seen in Figure 2, only the relative information from three neighboring vehicles are considered: the immediate lead and lag vehicles in the open lane and the immediate lead vehicle in the current lane.

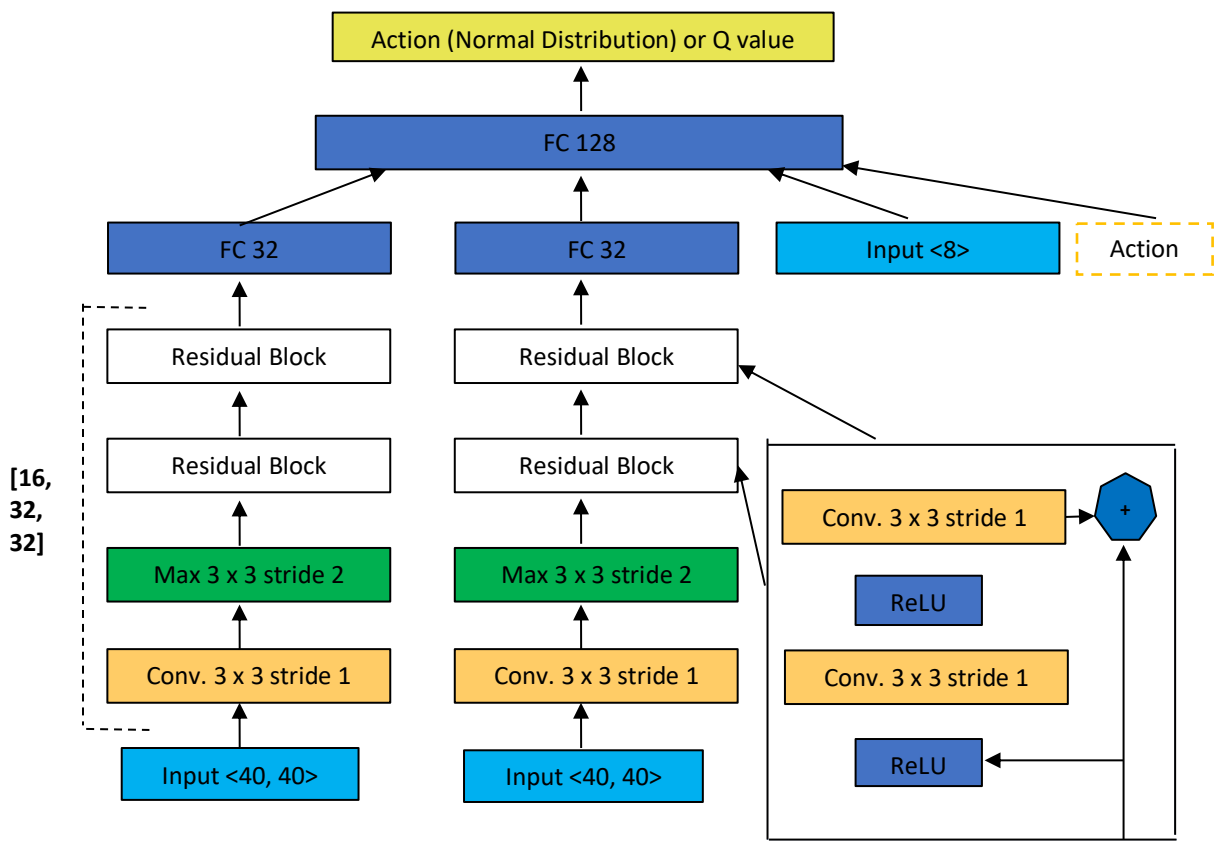

Figure 3 Convolutional Neural Network Architecture

The two grid maps give the subject vehicle a global view of the current traffic conditions in the work zone, while the 8-element vector is to provide the subject vehicle with more detailed local traffic information. In total, the proposed RL method takes 3,208 state variables. Given such a large input dimension, it is reasonable to use neural networks to capture the learned control policy.

*Neural Network Architecture*

When the state space is discrete and compact, the Q-function can be easily formulated as a table. However when state space is continuous and multi-dimensional, it is impossible to formulate the Q-function as a table or Monte Carlo Tree (*17*) such as in AlphaGo Zero (*18*). In such a case, the Q-function is often approximated by a parameterized function of states and actions $Q(s, a, w)$, and the learning process is to find the optimal parameter set $w$. This study adopts a modified impala convolutional neural network (*19*) to approximate the Q-function as well as the policy (actor) function.

As shown in Figure 3, the speed and acceleration grid maps are reshaped to two <40, 40> matrices and fed into a convolutional neural network (CNN). The CNN includes three main blocks with filter sizes 16, 32 and 32, respectively. The first two main blocks correspond to the speed and acceleration grid maps, and the last block is for the 8-element vector. Each of the first two main blocks starts from a 3*3 convolutional layer, includes a 3*3 maxpooling layer down sampling with stride 2, and serves two residual blocks which have a similar architecture as ResNet (*21*). The reason for adopting this CNN architecture is that as the network depth increases, accuracy gets saturated and degrades rapidly. The two residual blocks are included to increase the data sample efficiency by reusing activations from a previous layer until the adjacent layer learns its weights. This significantly simplifies the network and reduces the number of layers in it.

Via each of the first two main blocks, the raw state input is converted to a 32-dimension embedding which saves nonlinear and highly correlated information of the input. The two embeddings are concatenated with the 8-dimension vector as the input for the policy function, which output a final normal distribution with mean value and variance. The same CNN architecture is used for the Q function $Q(s, a)$ and value function $V(s)$. For the Q function, the action set is also added to the above concatenated vector to generate Q values for each state and action pair.

*Soft Actor Critic (SAC)*

On-policy RL algorithms such as Proximal Policy Optimization (PPO) (*23*), Asynchronous Actor Critic Agents (A3C) (*22*) and Trust Region Policy Optimization (TRPO) (*24*), although very popular, suffer from sample inefficiency because they need to generate new samples after each policy update and cannot utilize historical samples. On the contrary, Q-learning based off-policy approaches such as DDPG and DQN are able to learn efficiently from past experience sampled from memory replay buffer. However, these off-policy optimization algorithms are very sensitive to hyperparameters and require a lot of tuning to get the model converge. To address this issue, this study uses a novel Soft Actor Critic (SAC) RL. SAC is also an off-policy algorithm but includes new features to overcome the convergence brittleness problem.

The main difference between SAC and other off-policy RL algorithms is that SAC seeks to maximize not only the long-term rewards, but also the entropy of policy. It encourages policy exploration by assigning approximately the same probabilities to actions that have the same or similar Q-values. This feature prevents the policy from repeatedly selecting a small set of actions with high Q-values in the training process, while missing the chance of exploring other low Q-value actions that are potentially very rewarding in the long run. By encouraging policy exploration, SAC is able to address other off-policy algorithms' convergence problem.

$$J(\theta) = \sum_{t=1}^{T} E_{(s_t, a_t) \sim \rho_{\pi_\theta}} \left[ r(s_t, a_t) + \alpha \mathcal{H}\big(\pi_\theta(.\,|s_t)\big) \right] \tag{1}$$

The policy function is obtained by maximizing the objective function in Equation (1), which consists of a reward term and an entropy term $\mathcal{H}$ weighted by $\alpha$. SAC has three networks: a policy function $\pi$ parameterized by $\Phi$, a soft Q-approximator function $Q$ parameterized by $\theta$ and a state value function $V$ parameterized by $\psi$. The two separate approximators for $V$ and $Q$ functions are helpful for the learning process to converge.

To train the three CNNs, a series of loss functions are defined. The Policy network $\pi$ is trained by minimizing the following loss function in Equation (2):

$$\begin{aligned} \pi_{new} &= \arg\min_{\pi' \in \Pi} D_{KL}\left(\pi'(\cdot\,|s_t) \Big| \frac{\exp(Q^{\pi_{old}}(s_t,.))}{Z^{\pi_{old}}(s_t)}\right) \\ &= \arg\min_{\pi' \in \Pi} D_{KL}\left(\pi'(.\,|s_t) \Big| \exp\big(Q^{\pi_{old}}(s_t,.) - \log Z^{\pi_{old}}(s_t)\big)\right) \end{aligned} \tag{2}$$

To update the policy network, SAC restricts the policy to a subset of policies $\Pi$ which could be represented as a Gaussian distribution. In Equation (2), SAC uses the information projection defined in terms of the Kullback-Leibler divergence (*20*) between the old policy distribution and exponential of the old Q approximator function divided by the partition function $Z$ which normalizes the old Q distribution. Function $Z$ can be dropped since it is intractable in general and it does not affect the gradient with respect to the new policy.

Based on the Bellman equation, the soft Q-value can be computed iteratively starting from any function $Q: S \times A \rightarrow R$ given by Equation (3).

$$Q(s_t, a_t) = r(s_t, a_t) + \gamma E_{s_{t+1} \sim \rho_\pi(s)}[V(s_{t+1})] \tag{3}$$

where,

$$V(s_t) = E_{a_t \sim \pi}[Q(s_t, a_t) - \alpha \log \pi\,(a_t|s_t)] \tag{4}$$

$V(S_t)$ in Equation (4) is the soft state value function. The soft state value function is trained by minimizing the squared residual error in Equation (5).

$$J_V(\psi) = E_{s_t \sim \mathbb{D}}\left[\frac{1}{2}\big(V_\psi(s_t) - E[Q_\theta(s_t, a_t) - \log \pi_\Phi\,(a_t|s_t)]\big)^2\right] \tag{5}$$

with gradient,

$$\nabla_\psi J_V(\psi) = \nabla_\psi V_\psi(s_t)\big(V_\psi(s_t) - Q_w(s_t, a_t) + \log \pi_\theta\,(a_t|s_t)\big) \tag{6}$$

where $D$ is the distribution of previously sampled states and actions saved in the replay buffer. The soft Q function is trained by minimizing the soft Bellman residual (Equation (7)) using the stochastic gradient descent method.

$$J_Q(\theta) = E_{(s_t, a_t) \sim \mathbb{D}} \left[ \frac{1}{2} \left( Q_\theta(s_t, a_t) - \left( r(s_t, a_t) + \gamma E_{s_{t+1} \sim \rho_\pi(s)} \left[ V_{\bar{\psi}}(s_{t+1}) \right] \right) \right)^2 \right] \qquad (7)$$

with gradient,

$$\nabla_w J_Q(w) = \nabla_w Q_w(s_t, a_t) \left( Q_w(s_t, a_t) - r(s_t, a_t) - \gamma V_{\bar{\psi}}(s_{t+1}) \right) \qquad (8)$$

The target state value network $V_{\bar{\psi}}$ weights is updated by an exponential moving average considering the current value state network weights.

*Reward Shaping*

The main goal of reward shaping is to avoid creating "stop and go" traffic when a vehicle merges from the closed lane into the open lane. It requires the subject vehicle to keep a minimum safe distance with its lead vehicle and lag vehicle in the target/open lane when making a lane change. When the subject vehicle merges into the open lane, all vehicles surrounding it are supposed to continue smoothly without having to accelerate or decelerate.

For vehicles in the closed lane trying to merge, they are either in a non-terminal state or the terminal state. Non-terminal state represents when a vehicle is in Metering Zone II and adjusting its position, while terminal state is when a vehicle successfully merges into the open lane. The terminal state reward is calculated by Equation (9).

$$R = - \max(0, (70 - a_v) * 0.2) \qquad (9)$$

where $a_v$ is the average speed of all vehicles currently in Metering Zone II. The reward is negative if the subject vehicle is slower than $a_v$, since this may create a backward shockwave. In addition, if $dx_1 > v_1 * th_{min}$, $dx_2 > v_2 * th_{min}$ and $|(v_s - (v_1 + v_2)/2)| < 2$, $R += 10$, where $th_{min}$ is the minimum time headway, $v$ is for speed, $dx_1$ is the distance headway between the subject vehicle and the lag vehicle in the target lane, and $dx_2$ is the distance headway between the subject vehicle and the lead vehicle in the target lane.

For non-terminal states, the reward is determined based on the following equations:

$$R = -0.01 * acc^2 \qquad (10)$$
$$R -= 10 \text{ if a crash occurs} \qquad (11)$$
$$R -= 2.5 \text{ if } v_s < 30\text{km/h} \parallel v_s > 100\text{km/h} \parallel \text{front headway} < 2\text{m} \qquad (12)$$
$$R -= (v_s - 80) * 0.01 \text{ if } v_s > 80\text{km/h} \qquad (13)$$
$$R -= (60 - v_s) * 0.01 \text{ if } v_s < 60\text{km/h} \qquad (14)$$

Equation (10) encourages the subject vehicle to drive smoothly with minimum acceleration/deceleration. Equation (11) means that the simulation will be terminated and restarted if a crash occurs. Equations (12), (13) and (14) aim to minimize the vehicle's speed fluctuations around the speed limit (assuming 70km/h in this study). The reward functions are carefully designed and help the subject vehicle learn how to follow the lead vehicle without crash, travel at a reasonable speed, and maintain a safe distance with both the lead and lag vehicles in the target lane.

## SIMULATION ANALYSIS

### Experiment Design

This research adopts a microscopic simulation tool VISSIM to evaluate the performance of the proposed RL control strategy and to compare it with early merge (EM), late merge (LM) and no control (base case) under two input traffic volumes: 1,600 vph and 2,000 vph. As shown in Figure 1, a work zone on a two-lane highway with the right lane closed is considered. For all simulations conducted, the percentage of heavy vehicles is set to 3%, and the speed limit is set as 70 km/h. For each merge control and input volume combination, the simulation is run 10 times with different random seeds. Each simulation run lasts 45 minutes with the first 15 minutes serving as the warm-up period.

### Overall Mobility Performance

Table 1: Performance comparison of different control strategies

| Performance Measure | Merge Control Strategy | | | |
|---|---|---|---|---|
| | Base Case | Early Merge (EM) | Late Merge (LM) | RL Control |
| | Volume Input 1,600 vph | | | |
| **Average Delay (s)** | 274.8 | 121.9 (-55%) | 64.8 (-76%) | 4.2 (-97%) |
| **Throughput (vph)** | 1343 | 1424 (6%) | 1517 (13%) | 1596 (19%) |
| **Mean travel time (s)** | 384.3 | 231.4 (-40%) | 174.3 (-55%) | 116.5 (-70%) |
| | Volume Input 2,000 vph | | | |
| **Average Delay (s)** | 561.6 | 374.6 (-33%) | 372.5 (-34%) | 28.4 (-94%) |
| **Throughput (vph)** | 1341 | 1436 (7%) | 1526 (14%) | 1979 (48%) |
| **Mean travel time (s)** | 671.1 | 484.0 (-28%) | 482.0 (-28%) | 140.8 (-79%) |

*Note: numbers in parenthesis are relative differences, which are calculated as (control case – base case)/ (base case)\*100%*

Table 1 shows the mobility performance for different control strategies. The typical capacity for a two-lane highway with one lane closed is about 1,340 vph (*25*). When the input volume is 1,600 vph (i.e., above the normal capacity), the RL control gives the best results for all performance measures, followed by LM and EM. Compared to EM and LM, the delay from RL control in this case is much smaller. The throughput generated by RL control is almost the same as the input, demonstrating its superior mobility performance. Not surprisingly, no control yields the worst results. The average throughput without any control is 1,343 vph, which is consistent with the capacity reported in (*25*).

When the input volume increases from 1,600 vph to 2,000 vph, even the average delay for RL control goes up significantly. However, the trend observed under the 1,600 vph input volume level still holds. For EM and LM, the percentage improvements in terms of average delay and mean travel time both drop significantly compared to at the 1,600 vph demand level, while the percentage improvements in terms of throughput stay approximately the same.

Overall, the results in Table 1 suggest that RL control significantly improves work zone safety and mobility compared with traditional control strategies like EM and LM. Under oversaturated condition (e.g., 2,000 vph), the performance differences between EM and LM become marginal, especially in terms of average delay and mean travel time. On the other hand, RL control performs the best under both congested and oversaturated conditions.

**Vehicle Trajectory Diagram**

To illustrate how RL control adjusts the positions of individual vehicles and the benefits of doing so, the trajectories of vehicles in a randomly selected time frame are plotted in Figure 4, in which green lines are trajectories for vehicles in the right (closed) lane and red lines are for vehicles in the left (open) lane. Under RL control all green lines eventually turn red in the merging zone, meaning vehicles in the closed lane are able to successfully merge into the open lane. While for no control, Figure 4 clearly shows that many vehicles in the closed lane have to stop and wait for an extended period of time before they can merge into the open lane.

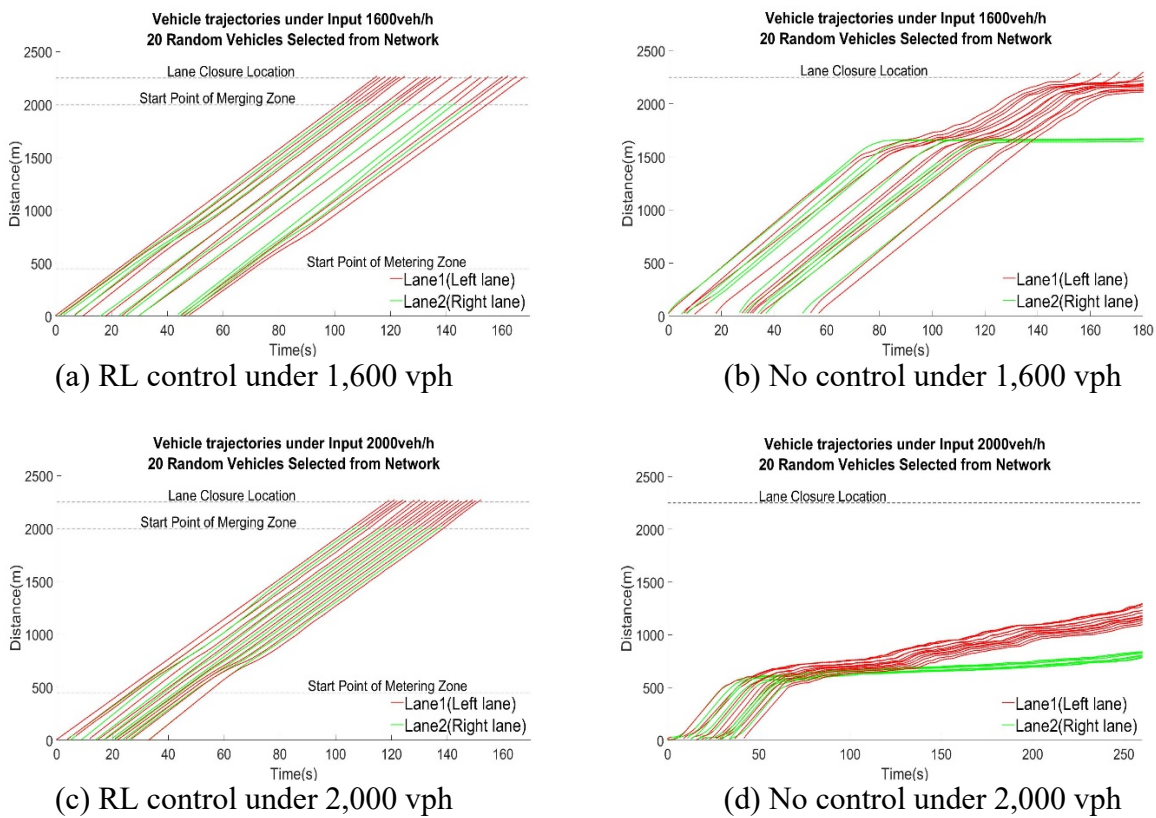

(a) RL control under 1,600 vph

(b) No control under 1,600 vph

(c) RL control under 2,000 vph

(d) No control under 2,000 vph

Figure 4 Vehicle Trajectory Diagrams.

Other than the mobility benefits of RL control clearly illustrated in Figure 4, the slopes of the trajectories show that RL control can help reduce rear-end crash risk, by avoiding sudden decelerations and stop-and-go traffic. Additionally, the no control trajectories show that some drivers in the closed lane have to wait for an extended amount of time to be able to merge and may become increasingly impatient. This intuitively may contribute to aggressive and unsafe behaviors such as forced merge, and increase the risk of angle crashes.

## Density

To further investigate how RL control performs, a VISSIM tool is developed to visualize how traffic density in the work zone changes over time and distance. The density maps for input volume = 1,600 vph under the RL control and LM strategies are presented in Figure 5, where the vertical axis is for time and the horizontal axis is for distance. A distance of 0 refers to the point 400 meters upstream of the metering zone. Larger distance values are for locations downstream of the origin. Also, red colors are for higher densities. Figure 5 clearly shows that compared to LM the RL control can better reduce and equalize the traffic densities of the open and closed lanes. Equal densities in both lanes can help reduce drivers' desire for lane changes (e.g., seeking higher speeds) and consequently angle crash risk. A smaller high-density area for RL control means the total vehicle time spent in stop-and-go traffic is less, suggesting that RL control is safer than LM at both input traffic volumes. Also, Figure 5 shows that the queues from the RL control grow at a much slower speed (i.e., backward forming shockwave speed) than the LM control. A slowly growing backward forming shockwave is likely to be less dangerous than a fast growing one.

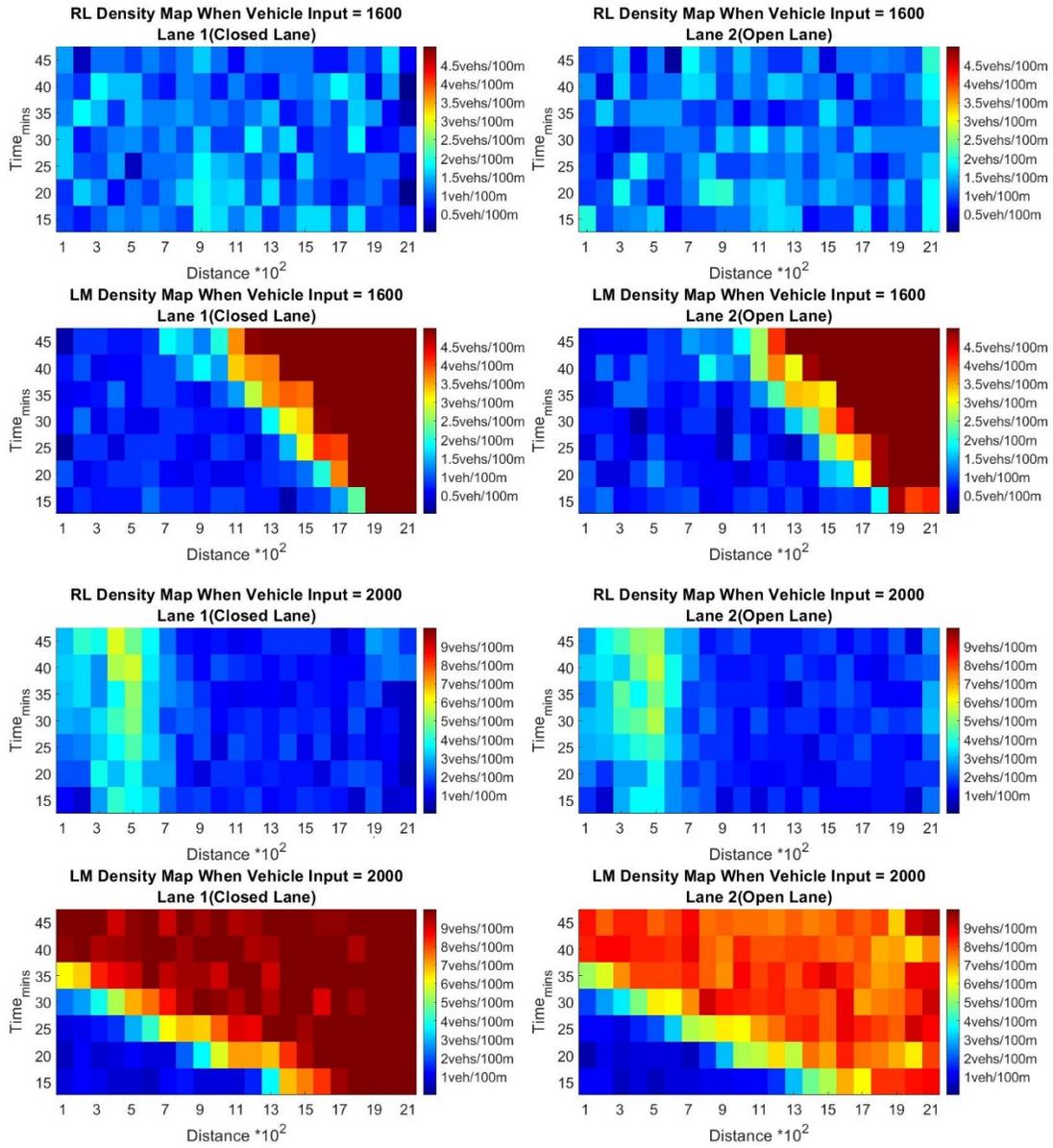

Figure 5 RL control and LM Density Map Comparison

## Acceleration and Distance Headway

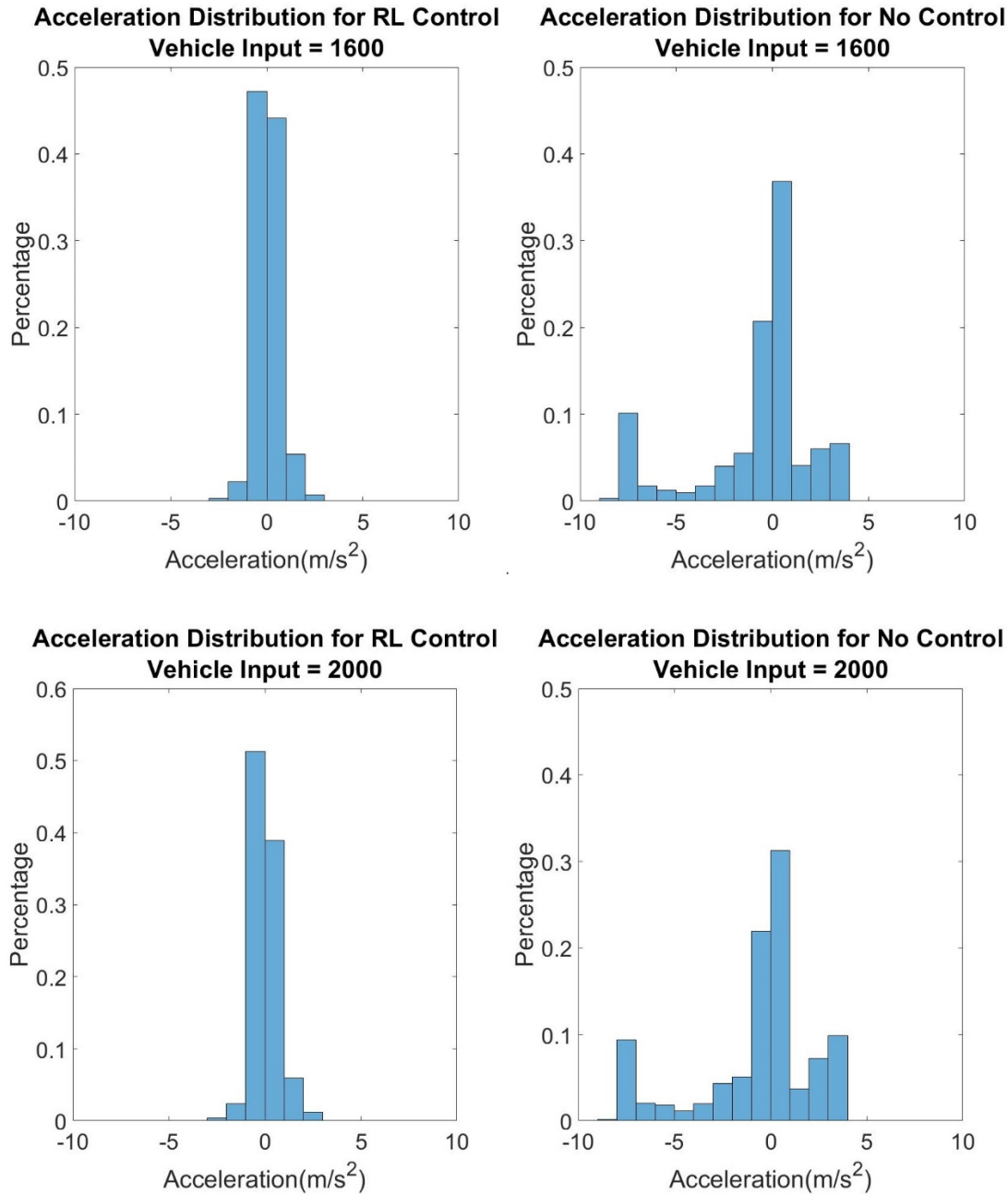

Figure 6 Acceleration Histogram.

A majority of crashes in highway work zones are rear-end crashes, which are often caused by sudden decelerations and stop-and-go traffic. Therefore, the stability of vehicle longitudinal acceleration behavior can be an important surrogate safety measure. Figure 6 shows the longitudinal acceleration distributions of vehicles under RL control and no control. The acceleration distributions for no control clearly are more spread out than those for RL control, and RL control generates much less sudden decelerations (e.g., <= -5 m²/s). This suggests that RL control is safer than no control and leads to smoother and more stable traffic flow.

The distance headway distributions for RL control and no control are also compared. Under both input flow conditions, overall RL control results in larger (safer) distance headways than no control in the merging zone.

## CONCLUSIONS AND DISCUSSION

This study proposes a cooperative highway work zone merge control strategy based on Soft Actor-Critic (SAC) reinforcement learning. This strategy is evaluated using VISSIM microscopic traffic simulation and compared with no control, late merge and early merge. The RL-based control performs significantly better than no control, early merge, and late merge under congested to extremely heavy traffic conditions in terms of both safety and mobility measures. Unlike other autonomous and connected vehicle control algorithms like CACC which increases the capacity of work zone by reducing vehicle time headway and reaction time, this RL-based control introduces two metering zones where vehicles adjust their positions relative to neighboring vehicles in the adjacent lane to achieve a collaborative and smooth merge and to maintain a safe time headway in the merging zone. The results also suggest the importance for automated vehicles to collaborate with each other in order to improve the overall system operations.

The proposed RL-based control strategy is applied to a two-lane highway work zone example. It can be further modified for multi-lane (more than two) highway work zones. For future studies, it would be interesting to investigate how to improve the system so that it can work in an environment with both automated and human-driven vehicles.

## AUTHOR CONTRIBUTION STATEMENT

The authors confirm contribution to the paper as follows: study conception and design: Tianzhu Ren, Yuanchang Xie; data collection: Tianzhu Ren; analysis and interpretation of results: Tianzhu Ren, Yuanchang Xie, Liming Jiang; draft manuscript preparation: Tianzhu Ren, Yuanchang Xie, Liming Jiang. All authors reviewed the results and approved the final version of the manuscript.